\documentclass[12pt,journal,onecolumn]{IEEEtran}
\IEEEoverridecommandlockouts

\usepackage{epsfig,rotating,setspace,latexsym,amsmath,epsf,amssymb,amsfonts,bm,theorem,cite,algorithm,graphicx,epsf,authblk,epstopdf,color,algpseudocode,bbm}

\newtheorem{corollary}{Corollary}
\newtheorem{definition}{Definition}
\newtheorem{theorem}{Theorem}
\newtheorem{lemma}{Lemma}
\newenvironment{Proof}[1]{\medskip\par\noindent{\bf Proof:\,}\,#1}{{\mbox{\,$\blacksquare$}\par}}

\allowdisplaybreaks

\begin{document}

\title{Timely Updates in Energy Harvesting Two-hop Networks: Offline and Online Policies\thanks{Ahmed Arafa was with the Department of Electrical and Computer Engineering at the University of Maryland College Park, when the majority of this work was done; he is now with the Electrical Engineering Department at Princeton University. Email: \emph{aarafa@princeton.edu}.}\thanks{Sennur Ulukus is with the Department of Electrical and Computer Engineering at the University of Maryland College Park. Email: \emph{ulukus@umd.edu}.}\thanks{This work was supported by NSF Grants CNS 13-14733, CCF 14-22111, CCF 14-22129, and CNS 15-26608; and has been presented in part at the IEEE Global Communications Conference, Singapore, December 2017.}}

\author{Ahmed Arafa,~\IEEEmembership{Member,~IEEE}, and Sennur Ulukus,~\IEEEmembership{Fellow,~IEEE}}

\maketitle

\begin{abstract}
A two-hop energy harvesting communication network is considered, in which measurement updates are transmitted by a source to a destination through an intermediate relay. Updates are to be sent in a {\it timely} fashion that minimizes the {\it age of information}, defined as the time elapsed since the most recent update at the destination was generated at the source. The source and the relay communicate using energy harvested from nature, which is stored in infinite-sized batteries. Both nodes use fixed transmission rates, and hence updates incur fixed delays (service times). Two problems are formulated: an {\it offline} problem, in which the energy arrival information is known a priori, and an {\it online} problem, in which such information is revealed causally over time. In both problems, it is shown that it is optimal to transmit updates from the source {\it just in time} as the relay is ready to forward them to the destination, making the source and the relay act as one combined node. A recurring theme in the optimal policy is that updates should be as {\it uniformly} spread out over time as possible, subject to energy causality and service time constraints. This is perfectly achieved in the offline setting, and is achieved almost surely in the online setting by a best effort policy.
\end{abstract}

\section{Introduction}

Developing low-latency communication policies is a crucial requirement for next generation communication networks, especially in applications pertaining to real-time status monitoring where information needs to be kept as {\it fresh} as possible at interested destinations. The {\it age of information} (AoI) metric has been recently introduced in the literature as a suitable performance metric to assess the freshness of data through basically measuring delays from receivers' perspectives. In applications where measurement updates regarding some phenomenon are to be sent to some destination over a period of time, the AoI is defined as the time elapsed since the most recent update at the destination was generated at the source. While it is clearly optimal to keep sending updates to maintain a low AoI, this might not be feasible all the time if transmitters have limited energy budgets. Therefore, managing the available energy becomes critical. In this paper, we consider an energy harvesting two-hop network, and propose {\it optimal} update policies that minimize the AoI in {\it offline} settings, where the energy arrival information is known prior to the start of communication, and {\it online} settings, where such information is only revealed causally over time.

Minimizing AoI has been extensively considered in the literature under various system settings and assumptions{. Interest in AoI started mainly in queuing systems, in which the early work of \cite{yates_age_1} shows that minimizing AoI is neither equivalent to minimizing transmission delay nor to maximizing throughput in a single server first-come first-serve queue with an infinite data buffer. In particular, it is optimal to design packet arrival rates in such systems to have an intermediate value between low arrival rates (that minimize transmission delay) and high arrival rates (that maximize throughput). One main reason behind this is that to minimize AoI one needs to keep the queue size relatively small, and at the same time keep delivering packets at a reasonable pace. This fundamental observation was then extended to the case of multiple sources in \cite{yates_age_mac}, and more recently in \cite{yates-age-mltpl-src}. References \cite{ephremides_age_random, ephremides_age_management, ephremides_age_non_linear} consider variations of the single source system and the queuing disciplines, such as randomly (out of order) arriving updates, update management and control, and nonlinear AoI metrics. Other variations considered include broadcasting, multicasting and multi streaming \cite{modiano-age-bc, soljanin-age-multicast, najm-age-multistream}. Reference \cite{sun-age-mdp} derives conditions on when it is optimal to submit new status updates right away after previous ones are delivered, and shows that, interestingly, it is not always optimal to do so; depending on the service time distribution it might be better to wait for a period of time before submitting the new update. Timely coding over erasure channels is studied in \cite{yates-age-erase-code}; connections between AoI and caching systems are drawn in \cite{yates-age-cache}; and those between AoI and remote estimation are drawn in \cite{sun-weiner}. Different from the single-hop settings considered in most works, references \cite{selen-age-gossip, atilla-age-multihop, shroff_age_multi_hop, modiano-age-multihop, shahab-age-multihop} focus on multihop settings. For instance, it is shown in \cite{shroff_age_multi_hop} that last-generated first-served policies are optimal in multihop networks under fairly arbitrary network topologies and packet arrival distributions. We refer the reader to the detailed survey in \cite{kosta-age-monograph} for more details on the various contexts in which AoI has been used.}

{Another setting in which AoI minimization has recently been studied is when transmitters/sensors rely on energy harvested from nature to communicate \cite{yates_age_eh, elif_age_eh, arafa-age-2hop, arafa-age-var-serv, elif-age-Emax, jing-age-online, jing-age-error-infinite-no-fb, jing-age-error-infinite-w-fb, baknina-age-coding, shahab-age-online-rndm, baknina-updt-info, arafa-age-sgl, arafa-age-rbr, arafa-age-online-finite, elif-age-online-threshold, arafa-age-erasure-no-fb, arafa-age-erasure-fb}. Energy harvesting offer the promise of providing energy self-sufficient and self-sustaining means of communications, and has been thoroughly studied in various contexts in the recent literature, see, e.g., \cite{ulukus_tutorial, qingqing-eh5g-overview}. In such settings, updates cannot be sent all the time, and optimal (AoI-minimal) energy management policies need to be carefully designed. The differentiating aspects between the works in \cite{yates_age_eh, elif_age_eh, arafa-age-2hop, arafa-age-var-serv, elif-age-Emax, jing-age-online, jing-age-error-infinite-no-fb, jing-age-error-infinite-w-fb, baknina-age-coding, shahab-age-online-rndm, baknina-updt-info, arafa-age-sgl, arafa-age-rbr, arafa-age-online-finite, elif-age-online-threshold, arafa-age-erasure-no-fb, arafa-age-erasure-fb} are mainly: battery capacity, offline/online knowledge of the energy arrivals, and service times (times for the updates to take effect). Aside from the conference version of this current work \cite{arafa-age-2hop}, which mainly discussed the offline setting, the above works focus on single-hop settings under both perfect and imperfect (noisy) channel conditions. A general conclusion from these works as a whole is that to minimize AoI, one needs to send status updates that are uniformly spread-out over time, to the extent allowed by the energy harvesting constraints. It is worth mentioning that the difference between AoI and delay minimization in energy harvesting communications has been discussed in \cite[Section~V]{arafa-age-var-serv}.}

In this paper, we characterize age-minimal status update policies in energy harvesting two-hop networks, in which a source is communicating with a destination through the help of an intermediate relay node. We consider the setting in which both the source and the relay are equipped with infinite batteries to save their incoming energy; offline and online knowledge of the energy arrivals; and {\it fixed} service times. In the offline scenario, the goal is to minimize the average AoI by a given deadline. We first solve a single-hop version of the problem, and then use its solution to solve for the two-hop version. Specifically, we show that it is optimal for the source to transmit a new update {\it just in time} as the relay is ready to forward it to the destination. Thus, the source and the relay can be treated as one combined node, after some appropriate transformations. The solution of the offline problem follows via an inter-update balancing algorithm, which aims at uniformly spreading the updates over time, up to the extent allowed by the energy arrivals and service times. In the online scenario, the goal is to minimize the long term average AoI. Energy arrives at the source and the relay according to two independent Poisson processes. We also show in this case that it is age-minimal to treat the source and the relay as one combined node. We then propose a {\it best effort} uniform update policy with service constraints, in which time is divided into slots of equal durations that depend on the service times, and an update is transmitted at the beginning of each time slot only if {\it both} the source and the relay have enough energy, otherwise both nodes stay silent and re-attempt transmission at the beginning of the next time slot. We prove that this policy is age-minimal by showing that it achieves a lower bound on the long term average AoI almost surely.

We note that the optimality of treating the source and the relay nodes as one combined node, and its consequences on not keeping any update packets waiting in the relay's data buffer, echoes the optimality results of last-generated first-served policies in the multihop setting considered in \cite{shroff_age_multi_hop}. We also note that such result is in contrast to the optimality of separable policies that maximize {\it throughput} in energy harvesting two-hop networks \cite{gunduz2hop, ruiZhangRelay}, where it is optimal to treat each node independently and send as many packets as possibly allowed by its own energy arrivals without considering those of the other node. In that sense, this work shows that if the metric considered is AoI, optimal policies are {\it inseparable}.

\section{System Model and Problem Formulation}

\begin{figure}[t]
\center
\includegraphics[scale=1]{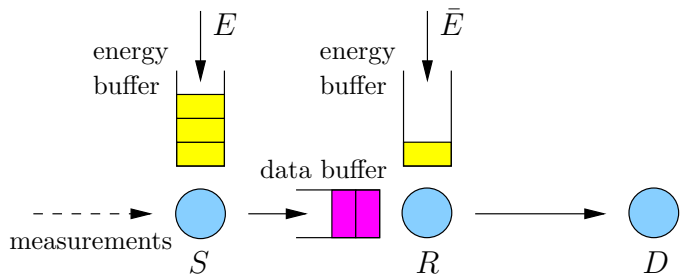}
\caption{Energy harvesting two-hop network. The source collects measurements and sends them to the destination through the relay.}
\label{fig_2hop_sys}
\end{figure}

A source node acquires measurement updates from some physical phenomenon and sends them to a destination, through the help of a half-duplex relay, see Fig.~\ref{fig_2hop_sys}. Both the source and the relay depend on energy harvested from nature to transmit their data, and are equipped with infinite-sized batteries to save their incoming energy. Energy arrives in packets of amounts $E$ and $\bar{E}$ at the source and the relay, respectively. Update packets are of equal length, and are transmitted using {\it fixed} rates at the source and the relay. We assume that one update transmission consumes one energy packet at a given node, and hence the number of updates at a given time is equal to the minimum of the number of energy packets that arrived at the source and the relay by that time. Under a fixed rate policy, each update takes $d$ and $\bar{d}$ amount of time to get through the source-relay channel and the relay-destination channel, respectively.\footnote{{We focus on error-free channels in this first look at multihop AoI minimization problems under energy harvesting constraints, in order to provide a clear understanding of the optimal status update policies' structure. We note that other works, e.g., \cite{jing-age-error-infinite-no-fb, jing-age-error-infinite-w-fb, arafa-age-erasure-no-fb, arafa-age-erasure-fb} have considered noisy channels situations for single-hop settings, in which updates are subject to erasures. Different from references \cite{jing-age-error-infinite-no-fb, jing-age-error-infinite-w-fb, arafa-age-erasure-no-fb, arafa-age-erasure-fb} though, which focus on zero (negligible) service times, we focus on non-zero, fixed, service times in the multihop setting of this work.}} 

The goal is to send updates as {\it timely} as possible, namely, such that the {\it age of information} (AoI) is minimized. The AoI metric at time $t$ is defined as
\begin{align}
a(t)\triangleq t-u(t)
\end{align}
where $u(t)$ is the time stamp of the latest received update packet at the destination before time $t$, i.e., the time at which it was acquired at the source.

{We note that the fixed energy consumption considerations in this work are derived by the fact that each status update typically consists of a value of the measurement acquired, and a time stamp denoting its acquisition time, that are both encoded into fixed-size packets of length, e.g., $B$ bits. The energy consumed to sense, encode, and transmit a single status update packet at the source is assumed to be equal to $E$, while the energy consumed to decode and forward a single status update packet at the relay is assumed to be equal to $\bar{E}$. Operating at such fixed energy consumption levels induces fixed transmission rate policies. For instance, if $g(e)$ is the rate in, e.g., bits per time units, used to send one status update packet with energy consumption $e$, then such update will take $B/g(e)$ time units of transmission delay to traverse through the communication channel. In our setting, such transmission delay is equal to $d$ and $\bar{d}$ at the source-relay channel and the relay-destination channel, respectively. The case in which transmitting devices are capable of using variable rates for different status updates has been previously considered in the single-hop setting in \cite{arafa-age-var-serv}. In this current work, however, we focus on fixed rates (and therefore fixed transmission delays) in order to fully characterize the optimal AoI minimizing solutions for multihop settings under both offline and online future energy arrival knowledge, which are formally introduced in the next subsections.}

\subsection{The Offline Setting}

In the offline setting, we consider a communication session of duration $T$ time units, in which energy arrival times are known a priori. Without loss of generality, we assume $a(0)=0$. The objective is to minimize the following quantity:
\begin{align}
A_T\triangleq\int_{0}^Ta(t)dt.
\end{align}

Source energy packets arrive at times $\{s_1,s_2,\dots,s_N\}\triangleq{\bm s}$, and relay energy packets arrive at times $\{\bar{s}_1,\bar{s}_2,\dots,\bar{s}_N\}\triangleq\bar{\bm s}$, where without loss of generality we assume that both the source and the relay receive $N$ energy packets, since each update consumes one energy packet in transmission from either node, and hence any extra energy arrivals at either the source or the relay cannot be used. Let $t_i$ and $\bar{t}_i$ denote the transmission time of the $i$th update at the source and the relay, respectively. We first impose the following constraints:
\begin{align}
t_i\geq s_i,~\bar{t}_i\geq\bar{s}_i,\quad 1\leq i\leq N, \label{eq_en_caus}
\end{align}
representing the {\it energy causality constraints} \cite{jingP2P} at the source and the relay, which mean that no energy packet can be used before being harvested. Next, we must have
\begin{align}
t_i+d\leq\bar{t}_i,\quad 1\leq i\leq N, \label{eq_data_caus}
\end{align}
to ensure that the relay does not forward an update before receiving it from the source, which represents the {\it data causality constraints} \cite{jingP2P}. We also have the {\it service time constraints}
\begin{align}
t_i+d\leq t_{i+1},~\bar{t}_i+\bar{d}\leq\bar{t}_{i+1},\quad1\leq i\leq N-1, \label{eq_1_tx}
\end{align}
which ensure that there can only be one transmission at a time at the source and the relay. Hence, $d$ and $\bar{d}$ represent the service (busy) time of the source and relay servers, respectively.

Transmission times at the source and the relay should also be related according to the half-duplex nature of the relay operation. For that, we must have the {\it half-duplex constraints}
\begin{align}
(t_i,t_i+d)\cap(\bar{t}_j,\bar{t}_j+\bar{d})=\emptyset,\quad \forall i,j, \label{eq_hf_dp_orig}
\end{align}
where $\emptyset$ denotes the empty set, since the relay cannot receive and transmit simultaneously. These constraints enforce that either the source transmits a new update after the relay finishes forwarding the prior one, i.e., $t_{i+1}\geq\bar{t}_i+\bar{d}$ for some $i$; or that the source delivers a new update before the relay starts transmitting the prior one, i.e., $t_{i+k}+d\leq\bar{t}_i$ for some $i$ and $k$. The latter case means that there are $k+1$ update packets waiting in the relay's data buffer just before time $\bar{t}_i$. We prove that this case is not age-optimal. To see this, consider the example of having $k+1=2$ update packets in the relay's data buffer waiting for service. The relay in this case has two choices at its upcoming transmission time: 1) forward the first update followed by the second one sometime later, or 2) forward the second update only and ignore the first one. These two choices yield different age evolution curves. We observe, geometrically, that $A_T$ under choice 2 is strictly less than that under choice 1. Since the source under choice 2 consumes an extra energy packet to send the first update unnecessarily, it should instead save this energy packet to send a new update after the first one is forwarded by the relay. Therefore, it is optimal to replace the half-duplex constraints in (\ref{eq_hf_dp_orig}) by the following reduced ones:
\begin{align}
\bar{t}_i+\bar{d}\leq t_{i+1},\quad 1\leq i\leq N-1. \label{eq_hf_dp}
\end{align}

Next, observe that (\ref{eq_1_tx}) can be removed from the constraints since it is implied by (\ref{eq_data_caus}) and (\ref{eq_hf_dp}). In conclusion, the constraints are now those in (\ref{eq_en_caus}), (\ref{eq_data_caus}), and (\ref{eq_hf_dp}). Finally, we add the following constraint to ensure reception of all updates by time $T$:
\begin{align}
\bar{t}_N+\bar{d}\leq T.
\end{align}

\begin{figure}[t]
\center
\includegraphics[scale=1]{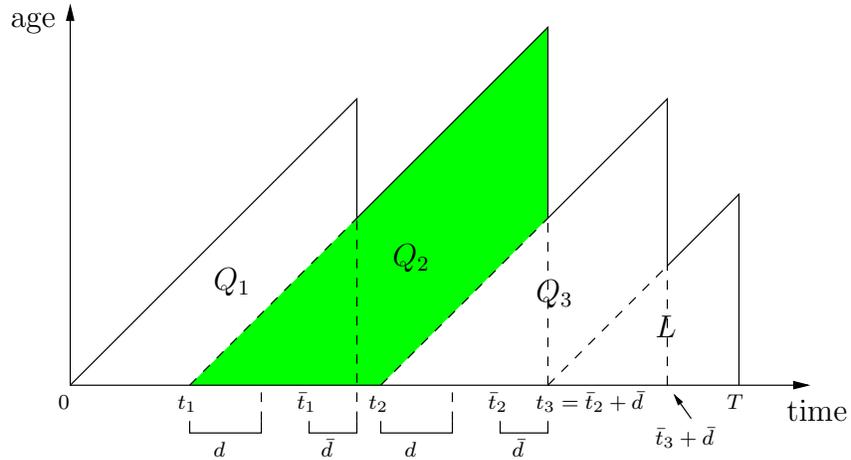}
\caption{Age evolution in a two-hop network with three updates.}
\label{fig_age_2hop}
\end{figure}

In Fig.~\ref{fig_age_2hop}, we present an example of the age of information in a system with 3 updates. The area under the curve representing $A_T$ is given by the sum of the areas of the trapezoids $Q_1$, $Q_2$, and $Q_3$, in addition to the area of the triangle $L$. The area of $Q_2$ for instance is given by $\frac{1}{2}\left(\bar{t}_2+\bar{d}-t_1\right)^2-\frac{1}{2}\left(\bar{t}_2+\bar{d}-t_2\right)^2$.
The objective is to choose feasible transmission times for the source and the relay such that $A_T$ is minimized. Computing the area under the age curve for general $N$ arrivals, we formulate the offline problem as follows:
\begin{align} \label{opt_main}
\min_{{\bm t},\bar{\bm t}}\quad&\sum_{i=1}^N\left(\bar{t}_i+\bar{d}-t_{i-1}\right)^2-\left(\bar{t}_i+\bar{d}-t_i\right)^2 + \left(T-t_N\right)^2 \nonumber \\
\mbox{s.t.}\quad&t_i\geq s_i,~\bar{t}_i\geq\bar{s}_i,\quad1\leq i\leq N \nonumber \\
&t_i+d\leq\bar{t}_i,\quad1\leq i\leq N\nonumber\\
&\bar{t}_i+\bar{d}\leq t_{i+1},\quad1\leq i\leq N.
\end{align}
with $t_0\triangleq0$ and $t_{N+1}\triangleq T$.

We note that the energy arrival times ${\bm s}$ and $\bar{\bm s}$, the transmission delays $d$ and $\bar{d}$, the session time $T$, and the number of energy arrivals $N$, are such that problem (\ref{opt_main}) has a feasible solution. This is true only if
\begin{align}
T&\geq\bar{s}_i+\left(N-i+1\right)\bar{d},\quad\forall i, \label{eq_feas_rl}\\
T&\geq s_i+\left(N-i+1\right)\left(d+\bar{d}\right),\quad\forall i, \label{eq_feas_s}
\end{align}
where (\ref{eq_feas_rl}) (resp. (\ref{eq_feas_s})) ensures that the $i$th energy arrival time at the relay (resp. source) is small enough to allow the reception of the upcoming $N-i$ updates within time $T$. 

\subsection{The Online Setting}

In the online setting, future energy arrival times are not known a priori; they get revealed causally over time. We assume, however, that there is some central controller that observes both energy arrival processes at the source and the relay, and takes decisions based on these simultaneous observations\footnote{A completely decentralized system model in which the source takes scheduling decisions independently from the relay is a possible direction for future work. In this work, we focus on the centralized problem.}. The energy arrival processes at the source and the relay are modeled as two independent Poisson processes of unit rate. The objective is to minimize the long term average area under the AoI curve. That is, to minimize
\begin{align}
\limsup_{T\rightarrow\infty}\frac{1}{T}\mathbb{E}\left[A_T\right].
\end{align}
Let us denote by $\mathcal{E}(t)$ and $\bar{\mathcal{E}}(t)$ the energy available in the source's and relay's batteries, respectively, right before time $t$, i.e., at time $t^-$. Therefore, the batteries evolve as follows:
\begin{align}
\mathcal{E}(t_i)&=\mathcal{E}(t_{i-1})-E+E\cdot\mathcal{A}\left([t_{i-1},t_i)\right), \label{eq_on_btry_src} \\
\bar{\mathcal{E}}\left(\bar{t}_i\right)&=\mathcal{E}\left(\bar{t}_{i-1}\right)-\bar{E}+\bar{E}\cdot\bar{\mathcal{A}}\left([\bar{t}_{i-1},\bar{t}_i)\right), \label{eq_on_btry_rly}
\end{align}
where $\mathcal{A}\left([a,b)\right)$ and $\bar{\mathcal{A}}\left([a,b)\right)$ denote the total number of energy arrivals in the interval $[a,b)$ at the source and the relay, respectively, which are two independent Poisson random variables with parameter $b-a$. We now write the energy causality constraints slightly differently from the offline setting as follows:
\begin{align}
\mathcal{E}(t_i)\geq E,~\bar{\mathcal{E}}\left(\bar{t}_i\right)\geq \bar{E},\quad \forall i. \label{eq_on_en_caus}
\end{align}
Let the feasible set $\mathcal{F}$ be the set of transmission times $\{t_i\}$ and $\{\bar{t}_i\}$ that adhere to the constraints in (\ref{eq_data_caus}), (\ref{eq_hf_dp}) and (\ref{eq_on_btry_src})-(\ref{eq_on_en_caus}), in addition to the following at time $0$: $\mathcal{E}(0)=E$ and $\bar{\mathcal{E}}(0)=\bar{E}$. The online problem is now to characterize the following quantity:
\begin{align} \label{opt_on}
\rho\triangleq\min_{{\bm t},\bar{{\bm t}}\in\mathcal{F}}\limsup_{T\rightarrow\infty}\frac{1}{T}\mathbb{E}\left[A_T\right].
\end{align}

We discuss problems (\ref{opt_main}) and (\ref{opt_on}) in details over the next two sections.

\section{The Offline Problem}

\subsection{Solution Building Block: The Single-User Channel}

\begin{figure}[t]
\center
\includegraphics[scale=1]{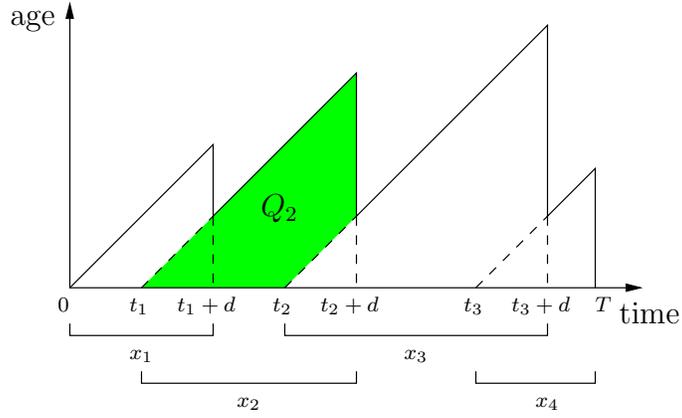}
\caption{Age evolution using in a single-user channel with three updates.}
\label{age_su}
\end{figure}

In this subsection, we solve the single-user version of problem (\ref{opt_main}); namely, when the source is communicating directly with the destination. We use the solution to the single-user problem in this subsection as a building block to solve problem (\ref{opt_main}) in the next subsection. In Fig.~\ref{age_su}, we show an example of the age evolution in a single-user setting. The area of $Q_2$ is now given by $\frac{1}{2}\left(t_2+d-t_1\right)^2-\frac{1}{2}d^2$. We compute the area under the age curve for general $N$ arrivals and formulate the single-user problem as follows:
\begin{align} \label{opt_su}
\min_{{\bm t}}\quad&\sum_{i=1}^N\left(t_i+d-t_{i-1}\right)^2+\left(T-t_N\right)^2\nonumber \\
\mbox{s.t.}\quad&t_i\geq s_i,\quad 1\leq i\leq N \nonumber \\
&t_i+d\leq t_{i+1},\quad 1\leq i\leq N,
\end{align}
where the second constraints are the service time constraints.

We note that reference \cite{elif_age_eh} considered problem (\ref{opt_su}) when the transmission delay $d=0$. We extend their results for a positive delay (and hence a finite transmission rate) in this subsection. We first introduce the following change of variables: $x_1\triangleq t_1+d$; $x_i\triangleq t_i-t_{i-1}+d,~2\leq i\leq N$; and $x_{N+1}\triangleq T-t_N$. These variables must satisfy $\sum_{i=1}^{N+1}x_i=T+Nd$, which reflects the dependent relationship between the new variables $\{x_i\}$. This can also be seen from Fig.~\ref{age_su}. Substituting by $\{x_i\}$ in problem (\ref{opt_su}), we get the following equivalent problem:
\begin{align} \label{opt_su_eq}
\min_{{\bm x}}\quad&\sum_{i=1}^{N+1}x_i^2 \nonumber\\
\mbox{s.t.}\quad&\sum_{i=1}^kx_i\geq s_k+kd,\quad1\leq k\leq N \nonumber\\
&x_i\geq2d,\quad2\leq i\leq N \nonumber\\
&x_{N+1}\geq d \nonumber\\
&\sum_{i=1}^{N+1}x_i=T+Nd.
\end{align}
Observe that problem (\ref{opt_su_eq}) is a convex problem that can be solved by standard techniques \cite{boyd}. For instance, we introduce the following Lagrangian:
\begin{align}
\mathcal{L}=&\sum_{i=1}^{N+1}x_i^2-\sum_{k=1}^N\lambda_k\left(\sum_{i=1}^kx_i- s_k-kd\right)-\sum_{i=2}^N\eta_i\left(x_i-2d\right)-\eta_{N+1}\left(x_{N+1}-d\right)\nonumber\\
&+\nu\left(\sum_{i=1}^{N+1}x_i-T-Nd\right),
\end{align}
where $\{\lambda_1,\dots,\lambda_N,\eta_2\dots,\eta_{N+1},\nu\}$ are Lagrange multipliers, with $\lambda_i,\eta_i\geq0$ and $\nu\in\mathbb{R}$. Differentiating with respect to $x_i$ and equating to 0 we get the following KKT conditions:
\begin{align}
x_1&=\sum_{k=1}^N\lambda_k-\nu, \label{eq_kkt_1}\\
x_i&=\sum_{k=i}^N\lambda_k+\eta_i-\nu,\quad2\leq i\leq N, \label{eq_kkt_2}\\
x_{N+1}&=\eta_{N+1}-\nu, \label{eq_kkt_3}
\end{align}
along with complementary slackness conditions
\begin{align}
\lambda_k\left(\sum_{i=1}^kx_i- s_k-kd\right)&=0,\quad1\leq k\leq N, \label{eq_slk_1} \\
\eta_i\left(x_i-2d\right)&=0,\quad1\leq i\leq N, \label{eq_slk_2} \\
\eta_{N+1}(x_{N+1}-d)&=0. \label{eq_slk_3}
\end{align}

We now have the following lemmas characterizing $\{x_i^*\}$, the optimal solution of problem (\ref{opt_su_eq}). Lemmas~\ref{thm_x_i_dec} and \ref{thm_x_N_N1} show that the sequence $\{x_i^*\}_{i=2}^{N+1}$ is non-increasing, and derive necessary conditions for it to strictly decrease. On the other hand, Lemma~\ref{thm_x_12} shows that $x_1^*$ can be smaller or larger than $x_2^*$, and derives necessary conditions for the two cases.

\begin{lemma} \label{thm_x_i_dec}
For $2\leq i\leq N-1$, $x_i^*\geq x_{i+1}^*$. Furthermore, $x_i^*> x_{i+1}^*$ only if $\sum_{j=1}^ix_j^*=s_i+id$.
\end{lemma}

\begin{Proof}
We show this by contradiction. Assume that for some $i\in\{2,\dots,N-1\}$ we have $x_i^*<x_{i+1}^*$. By (\ref{eq_kkt_2}), this is equivalent to having $\lambda_i+\eta_i<\eta_{i+1}$, i.e., $\eta_{i+1}>0$, which implies by complementary slackness in (\ref{eq_slk_2}) that $x_{i+1}^*=2d$. This means that $x_i^*<2d$, i.e., infeasible. Therefore $x_i^*\geq x_{i+1}^*$ holds. This proves the first part of the lemma.

To show the second part, observe that since $x_i^*>x_{i+1}^*$ holds if and only if $\lambda_i+\eta_i>\eta_{i+1}$, then either $\lambda_i>0$ or $\eta_i>0$. If $\eta_i>0$, then by (\ref{eq_slk_2}) we must have $x_i^*=2d$, which renders $x_{i+1}^*<2d$, i.e., infeasible. Therefore, $\eta_i$ cannot be positive and we must have $\lambda_i>0$. By complementary slackness in (\ref{eq_slk_1}), this implies that $\sum_{j=1}^ix_j^*=s_i+id$.
\end{Proof}

\begin{lemma} \label{thm_x_12}
$x_1^*>x_2^*$ only if $x_1^*=s_1+d$; while $x_1^*<x_2^*$ only if $x_i^*=2d$, for $2\leq i\leq N$.
\end{lemma}

\begin{Proof}
The necessary condition for $x_1^*$ to be larger than $x_2^*$ can be shown using the same arguments as in the proof of the second part of Lemma~\ref{thm_x_i_dec}, and is omitted for brevity. Let us now assume that $x_1^*$ is smaller than $x_2^*$. By (\ref{eq_kkt_1}) and (\ref{eq_kkt_2}), this occurs if and only if $\eta_2>\lambda_1$, which implies that $x_2^*=2d$ by complementary slackness in (\ref{eq_slk_2}). Finally, by Lemma~\ref{thm_x_i_dec}, we know that $\{x_i^*\}_{i=2}^N$ is non-increasing; since they are all bounded below by $2d$, and $x_2^*=2d$, then they must all be equal to $2d$.
\end{Proof}

\begin{lemma} \label{thm_x_N_N1}
$x_N^*\geq x_{N+1}^*$. Furthermore, $x_N^*>x_{N+1}^*$ only if at least: 1) $\sum_{i=1}^Nx_i^*=s_N+Nd$, or 2) $x_N^*=2d$ occurs.
\end{lemma}

The proof of Lemma~\ref{thm_x_N_N1} is along the same lines of the proofs of the previous two lemmas and is omitted for brevity.

{Observe that Lemma~\ref{thm_x_i_dec} states that (intermediate) inter-update delays cannot be increasing, and can only decreases following an update that was transmitted right away after an energy unit is harvested. Lemma~\ref{thm_x_12} states that an exception holds for the second inter-update delay, which can only increase, with respect to the first one, only if all updates starting from the second one to the end are sent back-to-back. Finally, Lemma~\ref{thm_x_N_N1} adds another condition to that of Lemma~\ref{thm_x_i_dec} for the last inter-update delay to decrease, which is to send the last update back-to-back after the previous one.}

We will use the results of Lemmas~\ref{thm_x_i_dec}, \ref{thm_x_12}, and \ref{thm_x_N_N1} to derive the optimal solution of problem (\ref{opt_su_eq}). To do so, one has to consider the relationship between the parameters of the problem: $T$, $d$, and $N$. For instance, one expects that if the session time $T$ is much larger than the minimum inter-update time $d$, then the energy causality constraints will be binding while the constraints enforcing one update at a time will not be, and vice versa. We formalize this idea by considering two different cases as follows.

\subsubsection{$Nd\leq T<(N+1)d$}

We first note that $Nd$ is the least value that $T$ can have for problem (\ref{opt_su_eq}) to admit a feasible solution. In this case, the following theorem shows that the optimal solution is achieved by sending all updates back-to-back with the minimal inter-update time possible to allow the reception of all of them by the end of the relatively small session time $T$.

\begin{theorem} \label{thm_T_sml}
Let $Nd\leq T<(N+1)d$. Then, the optimal solution of problem (\ref{opt_su_eq}) is given by
\begin{align}
x_1^*&=\max\left\{\frac{T-(N-2)d}{2},\max_{1\leq k\leq N}\left\{s_k-\left(k-2\right)d\right\}\right\}, \label{eq_x1_T_sml}\\
x_i^*&=2d, \quad 2\leq i\leq N, \label{eq_xi_T_sml}\\
x_{N+1}^*&=T-(N-2)d-x_1^*. \label{eq_x_N1_T_sml}
\end{align}
\end{theorem}

\begin{Proof}
We first argue that if $x_1^*\geq x_2^*(\geq2d)$, then $\sum_{i=1}^{N+1}x_i^*\geq(2N+1)d$. The last constraint in problem (\ref{opt_su_eq}) then implies that $T\geq(N+1)d$, which is infeasible in this case. Therefore, we must have $x_1^*<x_2^*$. By Lemma~\ref{thm_x_12}, this occurs only if $x_i^*=2d$ for $2\leq i\leq N$. Hence, we set $x_{N+1}=T-(N-2)d-x_1$, and observe that problem (\ref{opt_su_eq}) in this case reduces to a problem in only one variable $x_1$ as follows:
\begin{align}
\min_{x_1}\quad &x_1^2+\left(T-(N-2)d-x_1\right)^2 \nonumber \\
\mbox{s.t.}\quad &\max_{1\leq k\leq N}\left\{s_k-(k-2)d\right\}\leq x_1\leq T-(N-1)d,
\end{align}
whose solution is given by projecting the critical point of the objective function onto the feasible interval since the problem is convex \cite{boyd}. This directly gives (\ref{eq_x1_T_sml}).
\end{Proof}

\subsubsection{$T\geq (N+1)d$} \label{sec_su_largeT}

In this case, we propose an algorithmic solution that is based on the necessary optimality conditions in Lemmas~\ref{thm_x_i_dec}, \ref{thm_x_12}, and \ref{thm_x_N_N1}. We first solve problem (\ref{opt_su_eq}) without considering the service time constraints, i.e., assuming that the set of constraints $\{x_i\geq 2d,~2\leq i\leq N;~x_{N+1}\geq d\}$ is not active. We then check if any of these abandoned constraints is not satisfied, and optimally alter the solution to make it feasible.

Let us denote by $\left(P^e\right)$ problem (\ref{opt_su_eq}) without the set of constraints $\{x_i\geq 2d,~2\leq i\leq N;~x_{N+1}\geq d\}$, i.e., considering only the energy causality constraints. We then introduce the following algorithm to solve problem $\left(P^e\right)$:

\begin{definition}[Inter-Update Balancing Algorithm] \label{def_alg_eq}
Start by computing
\begin{align}
i_1\triangleq\arg\max\left\{s_1,\frac{s_2}{2},\dots,\frac{s_N}{N},\frac{T-d}{N+1}\right\},
\end{align}
where the set is indexed as $\{1,\dots,N+1\}$, and then set
\begin{align}
{\bar{x}_1}=\dots={\bar{x}_{i_1}}=\max\left\{s_1,\frac{s_2}{2},\dots,\frac{s_N}{N},\frac{T-d}{N+1}\right\}+d.
\end{align}
If $i_1=N+1$ stop, else compute

\begin{align}
i_2\triangleq\arg\max\left\{s_{i_1+1}-s_{i_1},\frac{s_{i_1+2}-s_{i_1}}{2},\dots,\frac{s_N-s_{i_1}}{N-i_1},\frac{T-d-s_{i_1}}{N+1-i_1}\right\},
\end{align}
where the set is indexed as $\{i_1+1,\dots,N+1\}$, and then set
\begin{align}
{\bar{x}_{i_1+1}}=\dots={\bar{x}_{i_2}}=\max\left\{s_{i_1+1}-s_{i_1},\frac{s_{i_1+2}-s_{i_1}}{2},\dots,\frac{s_N-s_{i_1}}{N-i_1},\frac{T-d-s_{i_1}}{N+1-i_1}\right\}+d.
\end{align}
If $i_2=N+1$ stop, else continue with computing $i_3$ as above. The algorithm is guaranteed to stop since it will at most compute $i_{N+1}$ which is equal to $N+1$ by construction.
\end{definition}

\begin{figure}[t]
\center
\includegraphics[scale=1.2]{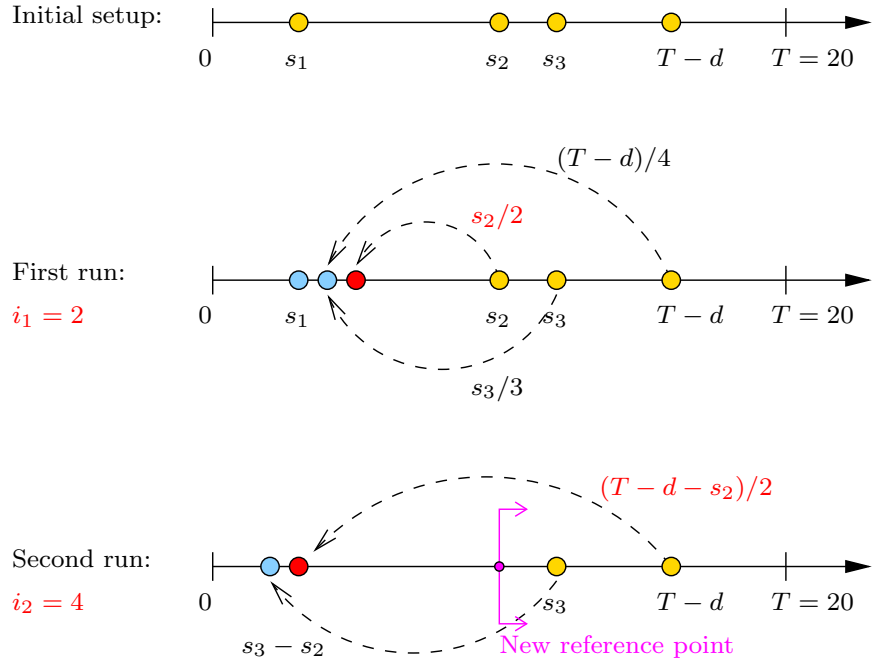}
\caption{{An example of how to apply the Inter-Update Balancing Algorithm in a system with $N=3$ energy arrivals at $s_1=3$, $s_2=10$, and $s_3=12$ time units. Service time $d=4$ time units, and the session ends at $T=20$ time units. The algorithm takes two runs to terminate. In each run, the maximum point highlighted in red is chosen by the algorithm.}}
\label{fig_alg_ex_1}
\end{figure}

Note that while computing $i_k$, if the $\arg\max$ is not unique, we pick the largest maximizer. Observe that the algorithm equalizes the $x_i$'s as much as allowed by the energy causality constraints. {In Fig.~\ref{fig_alg_ex_1}, we illustrate how the algorithm works using a graphical example. In the example, we have $N=3$ energy arrivals at $s_1=3$, $s_2=10$, and $s_3=12$ time units. Service time $d=4$ time units, and the session ends at $T=20$ time units. Applying the first run of the algorithm gives $i_1=2$. Therefore, we have $\bar{x}_1=\bar{x}_2=\frac{s_2}{2}+d=9$. Since $i_1=2<N+1=4$, the algorithm continues for a second run. Note that after the first run, the system has only one arrival left (at $s_3$), and it is as if the algorithm restarts again with a {\it new reference point} at $s_2$, as opposed to $0$. Applying the algorithm starting from $s_2$ in the second run gives $i_2=4$, which means $\bar{x}_3=\bar{x}_4=\frac{T-d-s_2}{N+1}+d=8$. Since $i_2=N+1=4$, the algorithm terminates.}

Let $\{\bar{x}_i\}_{i=1}^N$ be the output of the Inter-Update Balancing algorithm and let $\{x_i^e\}_{i=1}^N$ denote the optimal solution of problem $(P^e)$. We now have the following results:

\begin{lemma} \label{thm_eq_alg_prop}
$\{\bar{x}_i\}_{i=1}^N$ is a non-increasing sequence, and $\bar{x}_j>\bar{x}_{j+1}$ only if $\sum_{i=1}^j\bar{x}_i=s_j+jd$.
\end{lemma} 

\begin{lemma} \label{thm_eq_alg_opt}
$x_i^e=\bar{x}_i$, $1\leq i\leq N$.
\end{lemma}

The proofs of Lemmas~\ref{thm_eq_alg_prop} and \ref{thm_eq_alg_opt} are in Appendices~\ref{apndx_thm_eq_alg_prop} and \ref{apndx_thm_eq_alg_opt}, respectively.

We note that {Lemma~\ref{thm_eq_alg_prop} can be interpreted, operationally and in effect, as Lemma~\ref{thm_x_i_dec}. We also note that} Lemma~\ref{thm_eq_alg_opt} is similar to \cite[Theorem 1]{elif_age_eh}. In fact, the Inter-Update Balancing algorithm reduces to the optimal offline algorithm proposed in \cite{elif_age_eh} when $d=0$. When $d>0$, some change of parameters can still show the equivalence. The next corollary now follows.

\begin{corollary} \label{thm_partial_opt}
Consider problem $(P^e)$ with the additional constraint that $\sum_{i=1}^jx_i=s_j+jd$ holds for some $j\leq N$. Then, the optimal solution of the problem, under this condition, for time indices not larger than $j$ is given by $\{x_i^e\}_{i=1}^j$.
\end{corollary}

\begin{Proof}
This is direct by setting $T^\prime\triangleq s_j+d$ and $N^\prime\triangleq j-1$, and applying the Inter-Update Balancing algorithm on the problem with a reduced number of variables $\{x_1,\dots,x_{N^\prime+1}\}$.
\end{Proof}
 
The following theorem shows that the optimal solution of problem (\ref{opt_su_eq}), $\{x_i^*\}$, is found by equalizing the inter-update times as much as allowed by the energy causality constraints. If such equalization does not satisfy the minimal inter-update time constraints, we force it to be exactly equal to such minimum and adjust the last variable $x_{N+1}$ accordingly. The proof of the theorem is in Appendix~\ref{apndx_thm_amnd}.

\begin{theorem} \label{thm_amnd}
Let $T\geq(N+1)d$. If $x_i^e\geq2d,~2\leq i\leq N$ and $x_{N+1}^e\geq d$, then $x_i^*=x_i^e,~\forall i$. Else, let $n_0$ be the first time index at which $\{x_i^e\}$ is not feasible in problem (\ref{opt_su_eq}). Then, we have $n_0\leq N$. If $n_0>2$, we have
\begin{align}
x_i^*&=x_i^e,\quad1\leq i\leq n_0-1, \\
x_i^*&=2d,\quad n_0\leq i\leq N, \\
x_{N+1}^*&=T+Nd-\sum_{i=1}^Nx_i^*.
\end{align}
Otherwise, for $n_0=2$, $\{x_i^*\}$ is given by the above if $x_1^e=s_1+d$, else $\{x_i^*\}$ is given by (\ref{eq_x1_T_sml})-(\ref{eq_x_N1_T_sml}).
\end{theorem}

\subsection{Two-Hop Network: Solution of Problem (\ref{opt_main})}

We now discuss how to use the results of the single-user problem to solve problem (\ref{opt_main}). We have the following theorem:

\begin{theorem} \label{thm_2hop}
The optimal solution of problem (\ref{opt_main}) is given by the optimal solution of problem (\ref{opt_su}) after replacing $s_i$ by $\max\{\bar{s}_i,s_i+d\},~\forall i$; $d$ by $d+\bar{d}$; and $T$ by $T+d$.
\end{theorem}
\begin{Proof}
Let $f$ denote the objective function of problem (\ref{opt_main}). Differentiating $f$ with respect to $t_i$, $i\leq N-1$, we get $\frac{\partial f}{\partial t_i}=2\left(\bar{t}_i+\bar{d}-t_i\right)-2\left(\bar{t}_{i+1}+\bar{d}-t_i\right)$,
which is negative since $\bar{t}_{i+1}>\bar{t}_i$. We also have $\frac{\partial f}{\partial t_N}=2\left(\bar{t}_N+\bar{d}-t_N\right)-2\left(T-t_N\right)$, which is non-positive since $\bar{t}_N+\bar{d}\leq T$. Thus, $f$ is decreasing in $\{t_i\}_{i=1}^{N-1}$ and non-increasing in $t_N$. Therefore, the optimal $\{t_i^*\}$ satisfies the data causality constraints in (\ref{eq_data_caus}) with equality for all updates so as to be the largest possible and achieve the smallest $A_T$. Setting $t_i=\bar{t}_i-d,~\forall i$ in problem (\ref{opt_main}) we get
\begin{align}
f=\sum_{i=1}^N\left(\bar{t}_i+\bar{d}+d-\bar{t}_{i-1}\right)^2-N\left(\bar{d}+d\right)^2 + \left(T+d-\bar{t}_N\right)^2, 
\end{align}
with the constraints now being
\begin{align}
&\bar{t}_i\geq s_i+d,~\bar{t}_i\geq\bar{s}_i,\quad\forall i, \\
&\bar{t}_i+\bar{d}+d\leq\bar{t}_{i+1},\quad 1\leq i\leq N-1, \\
&\bar{t}_N+\bar{d}\leq T.
\end{align}
We now see that minimizing $f$ subject to the above constraints is exactly the same as solving problem (\ref{opt_su}) after applying the change of parameters mentioned in the theorem.
\end{Proof}

\begin{figure}[t]
\center
\includegraphics[scale=1]{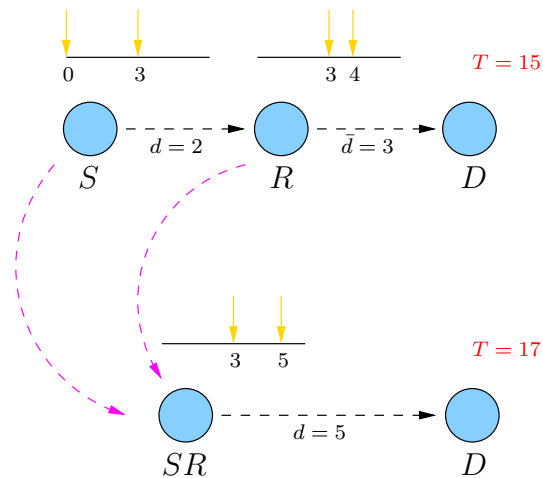}
\caption{{An example showing how to apply the results of Theorem~\ref{thm_2hop} going from the top, three-node system, to the bottom, two-node system, by combining the source and relay nodes, and applying the appropriate parameter conversions indicated in the theorem. Yellow arrows indicate energy arrivals.}}
\label{fig_th3_expln}
\end{figure}

Theorem~\ref{thm_2hop} shows that the source should send its updates {\it just in time} as the relay is ready to forward, and no update should wait for service in the relay's data buffer. Thus, the source and the relay act as one combined node that can send updates whenever it receives combined energy packets at times $\left\{\max\{\bar{s}_i,s_i+d\}\right\}$. {We show how to apply the theorem's premises on a graphical example in Fig.~\ref{fig_th3_expln}.} This fundamental observation can be generalized to multi-hop networks as well. Given $M>1$ relays, each node should send updates just in time as the following node is ready to forward, until reaching destination.

\section{The Online Problem}

In this section, we focus on characterizing the optimal solution of problem (\ref{opt_on}). We follow a two-step approach: we first derive a lower bound on the objective function of the problem, and then propose an online-feasible policy that achieves that lower bound, which shows its optimality. Let the random variable $N(T)$ denote the number of updates {\it received} by the destination within time $T$. We first begin by stating an upper bound on the update rate $N(T)/T$ in the limit as $T$ grows infinitely large in the following lemma:

\begin{lemma}
The following upper bound holds:
\begin{align} \label{eq_N_T_ub}
\limsup_{T\rightarrow\infty}\frac{N(T)}{T}\leq\min\left\{1,\frac{1}{d+\bar{d}}\right\} \quad \text{a.s.}
\end{align}
\end{lemma}

\begin{Proof}
By ignoring the energy causality constraints, and since an update can be received at the destination if and only if both the source and the relay have energy, we have
\begin{align}
N(T)\leq\min\left\{\sum_{i=1}^\infty\mathbbm{1}_{s_i\leq T}~,~\sum_{i=1}^\infty\mathbbm{1}_{\bar{s}_i\leq T}\right\} \quad \text{a.s.}
\end{align}
Since both energy arrival rates at the source and the relay are unit-valued, dividing both sides of the above equation by $T$ and taking limits shows that the update rate is upper-bounded by $1$. On the other hand, the service rate constraints imply that
\begin{align}
N(T)\leq\frac{T}{d+\bar{d}} \quad \text{a.s.},
\end{align}
which gives another upper bound of $1/\left(d+\bar{d}\right)$ on the update rate. Combining the two upper bounds gives the result in (\ref{eq_N_T_ub}).
\end{Proof}

The above lemma's result is quite intuitive. It shows that there can be two bottlenecks to the system's number of updates, one due to energy constraints, and the other due to service time constraints. Determining which one is in effect depends on the value of $d+\bar{d}$: if it is larger than $1$, which is the average energy arrival rate at both the source and the relay, then the system is more constrained by the service times; and if it is less than $1$, then the system is more constrained by the energy arrival rate. This simple observation leads to a variation in the optimal update policy as we show in the sequel. In the next lemma, we use the above result to derive a lower bound on the optimal solution, $\rho$, of problem (\ref{opt_on}). The proof of the lemma is in Appendix~\ref{apndx_thm_on_lb}.

\begin{lemma} \label{thm_on_lb}
The following lower bound holds:
\begin{align}
\rho\geq\max\left\{\frac{1}{2}+d+\bar{d}~,~\frac{3}{2}\left(d+\bar{d}\right)\right\}. \label{eq_on_lb}
\end{align}
\end{lemma}

Observe that Lemma~\ref{thm_on_lb} reemphasizes the fact that the system's bottleneck depends on the relationship between the total service time $d+\bar{d}$ and the average energy arrival rate. Clearly, if $d+\bar{d}<1$, the lower bound would be given by the first term in the maximum in (\ref{eq_on_lb}), and would be given by the second term otherwise. We will use this fact while devising the optimal policy below. Let us now define the following online update policy:

\begin{definition}[Best Effort Uniform Update Policy with Service Constraints]
The source schedules transmission of a new status update at times: $\ell_n\triangleq n\cdot\max\left\{1,d+\bar{d}\right\}$, $n=0,1,2,\dots$. At $\ell_n^-$, if the source and the relay have at least $E$ and $\bar{E}$ units of energy, respectively, then the update is transmitted from the source and gets forwarded by the relay directly once it is received. Otherwise, if either the source or the relay has an empty battery at $\ell_n^-$, then both nodes stay silent until the next scheduled transmission time $\ell_{n+1}$.
\end{definition}

Observe that the best effort uniform update policy with service constraints has a {\it scheduled} update rate of $1/\max\left\{1,d+\bar{d}\right\}=\min\left\{1,1/(d+\bar{d})\right\}$, which is the maximal possible update rate according to (\ref{eq_N_T_ub}). Our goal now is to show that such rate is indeed achievable. This is mainly shown by proving that the {\it failure} update rate, which is given by the ratio of the number of scheduled update times in which no transmission occurs (due to either the source or the relay having an empty battery) to the total elapsed time, is negligible in the long run. This is formally proven in the next theorem, the main result of this section, whose proof is in Appendix~\ref{apndx_thm_on_be}.

\begin{theorem} \label{thm_on_be}
The long term average AoI under the best effort uniform update policy with service constraints achieves the lower bound in Lemma~\ref{thm_on_lb}, and is therefore optimal.
\end{theorem}

We note that when $d+\bar{d}\geq1$, the optimal policy is equivalent to a {\it greedy} policy, in which an update is sent whenever both the source and the relay have enough energy. This is clear from the fact that the best effort uniform update policy with service constraints schedules updates every $d+\bar{d}$ time units, i.e., back-to-back. The performance of the greedy policy, however, deteriorates when $d+\bar{d}<1$. We touch upon this note again in the numerical results section below.

{We also note that, as in the offline setting, the optimal solution of the online setting is also extendible to general multihop scenarios with $M>1$ relays. This is attributed to the fact that the best effort uniform update policy with service constraints implicitly treats the source and the relay as one combined node, and only goes ahead with a transmission if {\it both} of them have enough energy simultaneously.}

\section{Numerical Results} \label{sec_num}

\begin{figure}[t]
\center
\includegraphics[scale=.65]{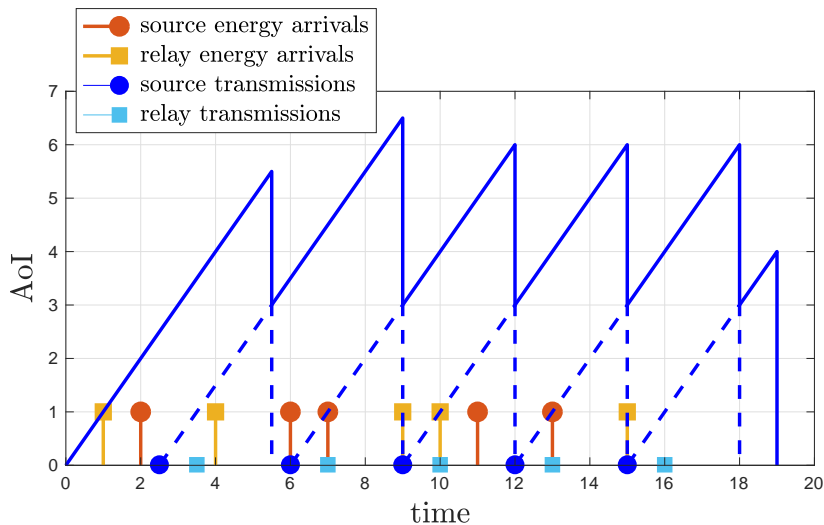}
\caption{{Optimal AoI curve achievable at the destination for the first offline example in Section~\ref{sec_num}. Filled circles (resp. squares) represent the source's (resp. relay's) energy arrivals/transmissions.}}
\label{fig_aoi_off_ex_1_2hop}
\end{figure}

We now present some numerical examples to further illustrate our results. We start by the offline ones. A two-hop network has energy arriving at times ${\bm s}=[2,6,7,11,13]$ at the source, and $\bar{{\bm s}}=[1,4,9,10,15]$ at the relay. A source transmission takes $d=1$ time unit to reach the relay; a relay transmission takes $\bar{d}=2$ time units to reach the destination. Session time is $T=19$. We apply the change of parameters in Theorem~\ref{thm_2hop} to get new energy arrival times ${\bm s}=[3,7,9,12,15]$, new transmission delay $d=3$, and new session time $T=20$. Then, we solve problem (\ref{opt_su_eq}) to get the optimal inter-update times, using the new parameters. Note that $T\geq(N+1)d=18$, whence the optimal solution is given by Theorem~\ref{thm_amnd}. We apply the Inter-Update Balancing algorithm to get ${\bm x}^e=[6.5,6.5,5.67,5.67,5.67,5]$. Hence, the first infeasible inter-update time occurs at $n_0=3$ ($x_3^e<2d=6$). Thus, we set: $x_1^*=x_1^e$ and $x_2^*=x_2^e$; $x_3^*=x_4^*=x_5^*=2d$; and $x_6^*=T+Nd-\sum_{i=1}^5x_i^*$. We see that ${\bm x}^*=[6.5,6.5,6,6,6,4]$ satisfies the conditions stated in Lemmas~\ref{thm_x_i_dec}, \ref{thm_x_12}, and \ref{thm_x_N_N1}. {In order to show how this translates back into the original two-hop network, in Fig.~\ref{fig_aoi_off_ex_1_2hop} we plot the achievable AoI curve at the destination, along with the optimal transmission times at both the source and the relay. We compute the optimal area under the AoI curve in this case to be $53.25$. We compare such area with the one achieved using an offline greedy policy, in which the source submits a new measurement whenever feasible, and the relay forwards it whenever possible. This is basically done by satisfying all the constraints of problem (\ref{opt_main}) with equality recursively as follows: $t_1=s_1$, $\bar{t}_1=\max\{\bar{s}_1,s_1+d\}$, $t_2=\max\{s_2,\bar{t}_1+\bar{d}\}$, $\bar{t}_2=\max\{\bar{s}_2,t_2+d\}$, and so on. The achievable area under the AoI curve under such offline greedy policy is $53.5$, which is strictly higher than that achieved by the optimal solution.}

We consider another example where energy arrives at times ${\bm s}=[0,4,4,9,13]$ and $\bar{{\bm s}}=[1,3,6,10,12]$, with $T=16$. Applying the change of parameters in Theorem~\ref{thm_2hop} we get $T=17<(N+1)d=18$, and hence we use the results of Theorem~\ref{thm_T_sml} to get ${\bm x}^*=[5,6,6,6,6,3]$. {The area under the AoI curve achieved with the optimal solution is given by $40$, while that achieved by the offline greedy policy is given by $42$.} We then increase $T$ to $18$. This is effectively $19$ according to Theorem~\ref{thm_2hop}, and therefore we apply Theorem~\ref{thm_amnd} results. The Inter-Update Balancing algorithm gives ${\bm x}^e=[5.8,5.8,5.8,5.8,5.8,5]$, and hence $n_0=2$. Since $x_1^e>s_1+d=4$, then the optimal solution is given by (\ref{eq_x1_T_sml})-(\ref{eq_x_N1_T_sml}) as ${\bm x}^*=[5,6,6,6,6,5]$. {As we compared before, the area under the AoI curve achieved with the optimal solution is given by $48$, while that achieved by the offline greedy policy is given by $50$.}

\begin{figure}[t]
\center
\includegraphics[scale=.65]{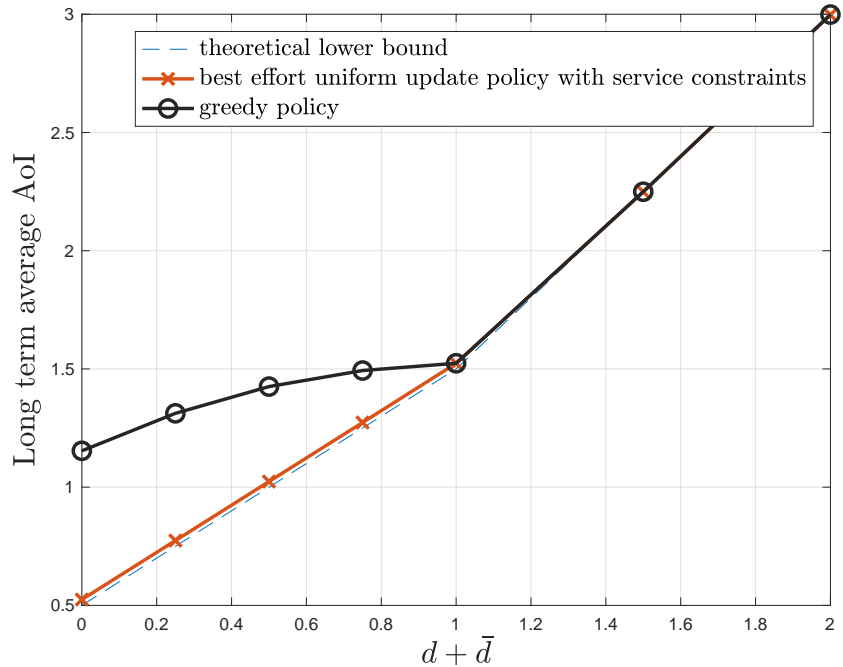}
\caption{Performance of the proposed best effort uniform updating policy with service constraints and a greedy policy versus the aggregate service time $d+\bar{d}$.}
\label{fig_online_results}
\end{figure}

We now present some online results. We create two independent sample paths according to unit-rate Poisson processes at the source and at the relay over the duration of $5\times10^3$ time units. We then use them to compute the long term average AoI achieved by the proposed best effort uniform updating policy with service constraints as a function of the aggregate service time $d+\bar{d}$. We also compare this to a greedy policy, in which an update is transmitted whenever both the source and the relay have enough energy, regardless of the value of $d+\bar{d}$. We compute an average performance of both policies over $10^3$ iterations, and plot the results in Fig.~\ref{fig_online_results}. In the figure, we also plot the theoretical lower bound of Lemma~\ref{thm_on_lb}. We see that the proposed policy and the lower bound are almost identical, as expected. We also see the superiority of the proposed policy in the low service time regime, i.e., when $d+\bar{d}<1$, compared to the greedy policy. In addition, as discussed after Theorem~\ref{thm_on_be}, the greedy policy becomes identical to the proposed policy in the high service time regime, when $d+\bar{d}\geq1$.

\begin{figure}
\center
\includegraphics[scale=.65]{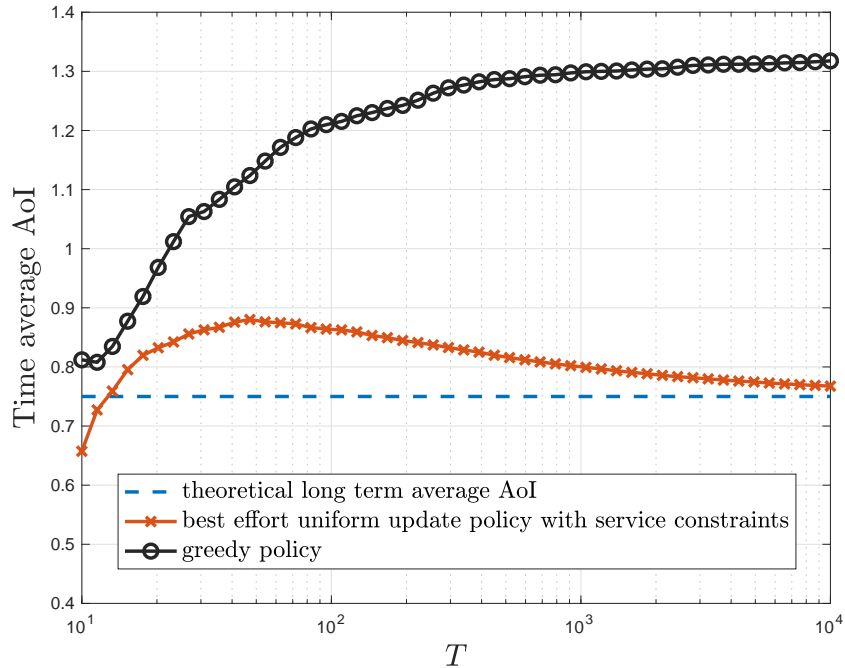}
\caption{{Time average AoI achieved under the best effort uniform update policy with service constraints and the greedy policy versus $T$ for a system in which $d+\bar{d}=0.25$ time units.}}
\label{fig_aoi_T_d_pt_25_smooth}
\end{figure}

{In Fig.~\ref{fig_aoi_T_d_pt_25_smooth}, we show how the time average AoI behaves as a function of $T$. We plot the achieved AoI under the best effort uniform update policy with service constraints and the greedy policy for values of $T$ ranging from $10$ to $10^4$ time units. In this example, $d+\bar{d}=0.25$ time units. We see from the figure that as $T$ grows, the greedy policy's performance deteriorates, while the performance of the best effort uniform update policy with service constraints approaches the theoretical long term average lower bound derived in Lemma~\ref{thm_on_lb}, confirming the results of Theorem~\ref{thm_on_be}. We note that the reason there is a small dip below the theoretical lower bound for small values of $T$ is that the bound in Lemma~\ref{thm_on_lb} is an asymptotic one that is effective only for relatively large values of $T$.}

\section{Conclusions and Future Directions}

Age-minimal transmission policies have been proposed for energy harvesting two-hop networks with fixed service times. It has been shown that the optimal update policy is such that the relay's data buffer should not contain any update packets waiting for service. Instead, updates should be transmitted from the source just in time as the relay is ready to forward them to the destination. In the offline setting, the optimal source transmission times, minimizing the average age of information by a given deadline, were found by an inter-update balancing algorithm that takes service times into consideration, as well as the knowledge of the incoming energy arrival information. In the online setting, a best effort uniform update policy with service constraints is shown to minimize the long term average age of information, in which time is uniformly divided into time slots of durations depending on service times, and an update is transmitted from the source at the beginning of each time slot only if both the source and the relay have enough energy.

The considered system model in the online setting of this paper is centralized, in which energy arrivals at both the source and the relay nodes are revealed {\it simultaneously} causally over time to some central controller that takes transmission decisions. It would be of interest to extend this work to a {\it decentralized} setting, in which both nodes take their own decisions independently of each other. Clearly, such decentralized problem is more challenging, but it may represent some practical applications where no central controller is present. Another line of extension would be to consider the finite battery case, in both centralized and decentralized scenarios, and extend the notions of threshold policies, which are optimal in single-hop settings \cite{arafa-age-online-finite, elif-age-online-threshold}, to multihop settings.

\appendix

\subsection{Proof of Lemma~\ref{thm_eq_alg_prop}} \label{apndx_thm_eq_alg_prop}

We show this by induction. Clearly, we have $\bar{x}_1=\bar{x}_2=\dots=\bar{x}_{i_1}=\frac{s_{i_1}}{i_1}+d$ by construction. Now assume that $\{\bar{x}_i\}_{i=1}^{i_k}$ is non-increasing, and consider $\{\bar{x}_i\}_{i=1}^{i_{k+1}}$. We know that $\bar{x}_{i_k+1}=\bar{x}_{i_k+2}=\dots=\bar{x}_{i_{k+1}}=\frac{s_{i_{k+1}}-s_{i_k}}{i_{k+1}-i_k}+d$ by construction. We now proceed by contradiction; assume that $\bar{x}_{i_{k+1}}>\bar{x}_{i_k}$. This means that the following holds:
\begin{align}
\frac{s_{i_{k+1}}-s_{i_k}}{i_{k+1}-i_k}>\frac{s_{i_k}-s_{i_{k-1}}}{i_k-i_{k-1}},
\end{align}
or equivalently
\begin{align} \label{eq_alg_dec_pf}
\frac{i_{k+1}-i_k}{i_{k+1}-i_{k-1}}s_{i_{k-1}}+\frac{i_k-i_{k-1}}{i_{k+1}-i_{k-1}}s_{i_{k+1}}>s_{i_k}.
\end{align}
Next, observe that the following holds by construction when choosing $x_{i_k}$:
\begin{align}
\frac{s_{i_k}-s_{i_{k-1}}}{i_k-i_{k-1}}\geq\frac{s_{i_{k+1}}-s_{i_{k-1}}}{i_{k+1}-i_{k-1}},
\end{align}
which is equivalent to
\begin{align}
\frac{i_{k+1}-i_k}{i_{k+1}-i_{k-1}}s_{i_{k-1}}+\frac{i_k-i_{k-1}}{i_{k+1}-i_{k-1}}s_{i_{k+1}}\leq s_{i_k}.
\end{align}
This contradicts (\ref{eq_alg_dec_pf}), and proves the first part of the lemma.

Now let us show the second part. Assume that $\bar{x}_{j+1}<\bar{x}_j$. Then necessarily we must have $j=i_k$ for some $i_k$, or else they should be equal. Therefore, by construction, we have
\begin{align} \label{eq_alg_when_dec}
\sum_{i=1}^{i_k}\bar{x}_i=&\left(i_k-i_{k-1}\right)\left(\frac{s_{i_k}-s_{i_{k-1}}}{i_k-i_{k-1}}+d\right)+\left(i_{k-1}-i_{k-2}\right)\left(\frac{s_{i_{k-1}}-s_{i_{k-2}}}{i_{k-1}-i_{k-2}}+d\right)+\dots \nonumber \\
&+i_1\left(\frac{s_{i_1}}{i_1}+d\right) \nonumber \\
=&s_{i_k}+i_kd.
\end{align}
This concludes the proof.

\subsection{Proof of Lemma~\ref{thm_eq_alg_opt}} \label{apndx_thm_eq_alg_opt}

Let $\{\bar{x}_i\}_{i=1}^N$ be the output of the Inter-Update Balancing algorithm and let $\{x_i^e\}_{i=1}^N$ denote the optimal solution of problem $(P^e)$. We first show that $\{\bar{x}_i\}_{i=1}^N$ is feasible. Let $j$ be such that $i_k<j\leq i_{k+1}$. Then
\begin{align}
\sum_{i=1}^j\bar{x}_i&=\sum_{i=1}^{i_k}\bar{x}_i+\sum_{i=i_k+1}^j\bar{x}_i \nonumber \\
&=s_{i_k}+i_kd+\left(j-i_k\right)\left(\frac{s_{i_{k+1}}-s_{i_k}}{i_{k+1}-i_k}+d\right) \label{eq_a} \\
&\geq s_{i_k}+jd+\left(j-i_k\right)\left(\frac{s_j-s_{i_k}}{j-i_k}\right) \label{eq_b} \\
&=s_j+jd,
\end{align}
where $(\ref{eq_a})$ follows by (\ref{eq_alg_when_dec}), and $(\ref{eq_b})$ follows since, by construction, we have
\begin{align}
\frac{s_{i_{k+1}}-s_{i_k}}{i_{k+1}-i_k}\geq\frac{s_j-s_{i_k}}{j-i_k},\quad\forall i_k<j\leq i_{k+1}.
\end{align}
Finally, note that the stopping criterion of the algorithm is when $i_L=N+1$ for some $i_L$. Whence, we have
\begin{align}
\sum_{i=1}^{N+1}\bar{x}_i=&i_1\left(\frac{s_{i_1}}{i_1}+d\right)+\left(i_2-i_1\right)\left(\frac{s_{i_2}-s_{i_1}}{i_2-i_1}+d\right)\!+\dots+\!\left(N+1-i_{L-1}\right)\left(\frac{T-d-s_{i_{L-1}}}{N+1-i_{L-1}}+d\right)\nonumber\\
=&(N+1)d+(T-d)=T+Nd.
\end{align}
This shows that that $\{\bar{x}_i\}_{i=1}^N$ is feasible.

Next, we show that $x_i^e=\bar{x}_i,~\forall i$. We show this by contradiction. Let $x_i^e=\bar{x}_i$ for $1\leq i\leq m-1$ and let $x_m^e\neq\bar{x}_m$, i.e., $m$ is the first time index at which the two sequences are different. We now consider two cases as follows.

First, assume $x_m^e<\bar{x}_m$. Note that it must be the case that $i_{k_1}<m\leq i_k$ for some $i_k$. Therefore, $\bar{x}_i=\bar{x}_m,~\forall m\leq i\leq i_k$, by construction. By (\ref{eq_kkt_2}), we have that $\{x_i^e\}$ is non-increasing since $\eta_i=0$ and $\lambda_i\geq0$. Therefore, $x_i^e<\bar{x}_i,~\forall m\leq i\leq i_k$, and hence $\sum_{i=1}^{i_k}x_i^e<\sum_{i=1}^{i_k}\bar{x}_i=s_{i_k}+i_kd$, i.e., the allegedly-optimal policy is not feasible. Therefore $x_m^e\geq \bar{x}_m$.

Second, assume $x_m^e>\bar{x}_m$. Since $\sum_{i=1}^{N+1}x_i^e=\sum_{i=1}^{N+1}\bar{x}_i$, therefore there must exist some time index $l>m$ such that $x_l^e<\bar{x}_l$. Now let $\epsilon\triangleq \min\{x_m^e-\bar{x}_m,\bar{x}_l-x_l^e\}$, and consider a new policy $\{\tilde{x}_i\}$ which is equal to $\{x_i^e\}$ except at time indices $m$ and $l$, with $\tilde{x}_m=x_m^e-\epsilon$ and $\tilde{x}_l=x_l^e+\epsilon$. Since $\tilde{x}_m\geq\bar{x}_m$, the new policy is feasible. In addition, by convexity of the square function, the following holds \cite{boyd}:
\begin{align}
\left(\tilde{x}_m\right)^2+\left(\tilde{x}_l\right)^2<\left(x_m^*\right)^2+\left(x_l^*\right)^2,
\end{align}
which means that the new policy achieves a lower age, rendering $\{x_i^e\}$ suboptimal.

The above arguments show that we must have $x_i^e=\bar{x}_i,~\forall i$. This completes the proof.

\subsection{Proof of Theorem~\ref{thm_amnd}} \label{apndx_thm_amnd}

The first part of the theorem follows directly since the solution of the less constrained problem $\left(P^e\right)$ is optimal if feasible in problem (\ref{opt_su_eq}). Next, we prove the second part.

We first show that $n_0\leq N$ by contradiction. Assume that $n_0=N+1$, i.e., $x_{N+1}^e<d$ and $x_N^e\geq 2d>x_{N+1}^e$. By Lemma~\ref{thm_eq_alg_prop}, this means that $\sum_{i=1}^Nx_i^e=s_N+Nd$. Hence, $x_{N+1}^e=T+Nd-s_N-Nd=T-s_N$, which cannot be less than $d$ by the feasibility assumption in (\ref{eq_feas_rl}). Thus, $n_0\leq N$.

Now let $n_0>2$ and observe that $x_{n_0}^e<2d\leq x_{n_0-1}$. Thus, by Lemma~\ref{thm_eq_alg_prop}, we must have $\sum_{i=1}^{n_0-1}x_i^e=s_{n_0-1}+(n_0-1)d$. Now let us show that the proposed policy is feasible; we only need to check whether $x_{N+1}^*\geq d$. Towards that, we have
\begin{align}
x_{N+1}^*&=T+Nd-\sum_{i=1}^{n_0-1}x_i^*-\left(N-n_0+1\right)2d \nonumber \\
&=T-s_{n_0-1}-(N-n_0+1)d \nonumber \\
&\geq d,
\end{align}
where the last inequality follows by the feasibility assumption in (\ref{eq_feas_rl}). Therefore, the proposed policy is feasible.

We now show that it is optimal as follows. Assume that there exists another policy $\{\tilde{x}_i\}$ that achieves a lower age than $\{x_i^*\}$. We now have two cases. First, assume that $\sum_{i=1}^{n_0-1}\tilde{x}_i=s_{n_0-1}+(n_0-1)d$. then by Corollary~\ref{thm_partial_opt} we must have $\tilde{x}_i=x_i^*$ for $1\leq i\leq n_0-1$. Now for $n_0\leq i\leq N$, if $\tilde{x}_i>x_i^*=2d$, this means that $\tilde{x}_{N+1}<x_{N+1}^*$ to satisfy the last constraint in (\ref{opt_su_eq}). Since $\sum_{i=n_0}^{N+1}\tilde{x}_i=\sum_{i=n_0}^{N+1}x_i^*$, then by convexity of the square function, $\sum_{i=n_0}^{N+1}\left(\tilde{x}_i\right)^2>\sum_{i=n_0}^{N+1}\left(x_i^*\right)^2$ \cite{boyd}, and hence $\{\tilde{x}_i\}$ cannot be optimal. Second, assume that $\sum_{i=1}^{n_0-1}\tilde{x}_i>s_{n_0-1}+(n_0-1)d=\sum_{i=1}^{n_0-1}x_i^*$. Since $\tilde{x}_i\geq x_i^*=2d$ for $n_0\leq i\leq  N$, and $\sum_{i=1}^{N+1}\tilde{x}_i=\sum_{i=1}^{N+1}x_i^*$, then we must have $\tilde{x}_{N+1}<x_{N+1}^*$. Thus, $\sum_{i=1}^{N+1}\left(\tilde{x}_i\right)^2>\sum_{i=1}^{N+1}\left(x_i^*\right)^2$ by convexity of the square function \cite{boyd}, and $\{\tilde{x}_i\}$ cannot be optimal.

Finally, let $n_0=2$. If $x_1^e=s_1+d$, then the proof follows by the arguments for the $n_0>2$ case. Else if $x_1^e>s_1+d$, then $x_1^e=x_2^e\geq x_{N+1}^e$ by Lemma~\ref{thm_eq_alg_prop}. Since $\{x_i^e\}_{i=2}^N$ have to increase to at least $2d$, then $x_1^e+x_{N+1}^e$ has to decrease to satisfy the last constraint in (\ref{opt_su_eq}). However, one cannot increase $x_1^e$ to $2d$ or more and compensate that by decreasing $x_{N+1}^e$, by convexity of the square function. Thus, $x_1^*<x_2^*$, and Lemma~\ref{thm_x_12} shows that the results of Theorem~\ref{thm_T_sml} follow to give (\ref{eq_x1_T_sml})-(\ref{eq_x_N1_T_sml}).

\subsection{Proof of Lemma~\ref{thm_on_lb}} \label{apndx_thm_on_lb}

We start by noting that, as observed in the proof of Theorem~\ref{thm_2hop}, the objective function is minimized if the data causality constraints hold with equality at all times. Ignoring the remainder of the constraints, we have that
\begin{align}
\rho&\geq\limsup_{T\rightarrow\infty}\frac{1}{T}\mathbb{E}\left[\sum_{i=1}^{N(T)}\frac{1}{2}\left(\bar{t}_i+d+\bar{d}-\bar{t}_{i-1}\right)^2 - \frac{N(T)}{2}\left(d+\bar{d}\right)^2 + \frac{1}{2}\left(T+d-\bar{t}_{N(T)}\right)^2\right] \label{eq_on_lb_1} \\
&\geq\mathbb{E}\left[\liminf_{T\rightarrow\infty}\frac{N(T)}{2T}\frac{1}{N(T)}\sum_{i=1}^{N(T)}\frac{1}{2}\left(\bar{t}_i+d+\bar{d}-\bar{t}_{i-1}\right)^2 - \limsup_{T\rightarrow\infty}\frac{N(T)}{2T}\left(d+\bar{d}\right)^2\right] \label{eq_on_lb_2} \\
&\geq\mathbb{E}\left[\liminf_{T\rightarrow\infty}\frac{N(T)}{2T}\left(\frac{1}{N(T)}\sum_{i=1}^{N(T)}\bar{t}_i+d+\bar{d}-\bar{t}_{i-1}\right)^2\right] - \frac{1}{2}\min\left\{1,\frac{1}{d+\bar{d}}\right\}\left(d+\bar{d}\right)^2 \label{eq_on_lb_3} \\
&=\mathbb{E}\left[\liminf_{T\rightarrow\infty}\frac{N(T)}{2T}\left(\frac{\bar{t}_{N(T)}}{N(T)}+d+\bar{d}\right)^2\right] - \frac{1}{2}\min\left\{\left(d+\bar{d}\right)^2,d+\bar{d}\right\} \\
&=\frac{1}{2}\min\left\{1,\frac{1}{d+\bar{d}}\right\}\left(\frac{1}{\min\left\{1,\frac{1}{d+\bar{d}}\right\}}+d+\bar{d}\right)^2 - \frac{1}{2}\min\left\{\left(d+\bar{d}\right)^2,d+\bar{d}\right\} \label{eq_on_lb_4} \\
&=\frac{1}{2}\frac{1}{\min\left\{1,\frac{1}{d+\bar{d}}\right\}}+d+\bar{d},
\end{align}
where (\ref{eq_on_lb_2}) follows by ignoring the last term in (\ref{eq_on_lb_1}) and using Fatou's lemma; (\ref{eq_on_lb_3}) follows by Jensen's inequality (convexity of the square function) and using the upper bound in (\ref{eq_N_T_ub}); and (\ref{eq_on_lb_4}) follows by considering a policy that achieves the bound in (\ref{eq_N_T_ub}). The result now follows after simple manipulations of the last equation above.

\subsection{Proof of Theorem~\ref{thm_on_be}} \label{apndx_thm_on_be}

We show this by analyzing the performance of a slightly different update policy that {\it cannot perform better} than the proposed best effort uniform update policy with service constraints, and proving that it achieves the lower bound in Lemma~\ref{thm_on_lb}. Such policy is different from the proposed policy in that whenever the source or the relay have an empty battery at some $\ell_n^-$, then not only they both stay silent until the next scheduled update time $\ell_{n+1}$, but whichever node has a non-zero amount of energy in its battery at $\ell_n^-$ {\it dumps} that amount. Despite the fact that one node dumping its available energy when the other node does not have any seems quite wasteful, it is considered mainly to facilitate the analysis. We now proceed with that dumping condition in action.

\begin{figure}[t]
\center
\includegraphics[scale=.8]{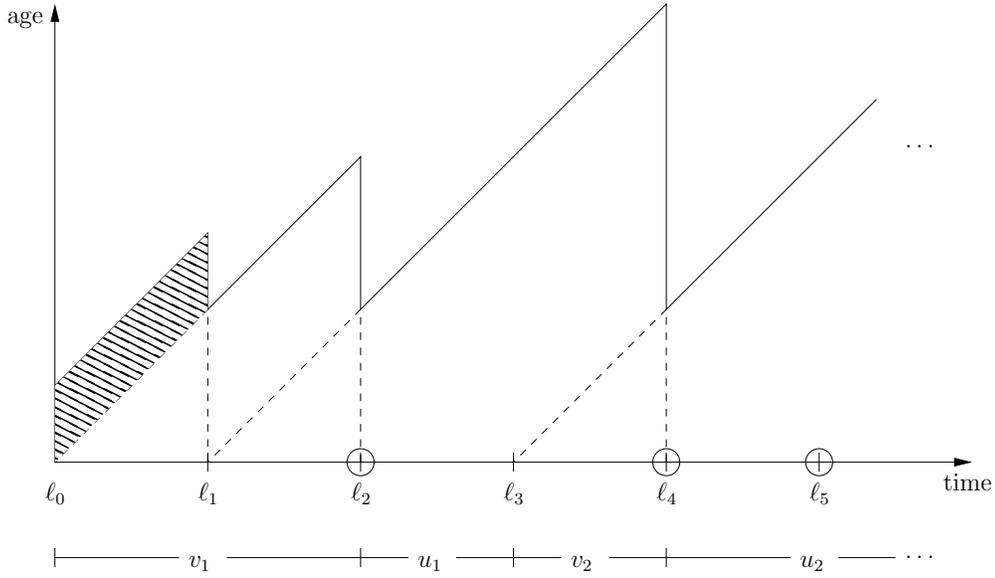}
\caption{An example of the age evolution versus time under the best effort uniform update policy with service constraints. $u_i$'s represent intervals during which scheduled updates fail to be transmitted due to lack of energy at either the source or the relay nodes, and $v_i$'s represent intervals during which scheduled updates are successfully transmitted. Circles denote failed scheduled update times.}
\label{fig_on_ex_u_v}
\end{figure}

We adopt the proof techniques involved in proving the optimality of best effort uniform updating policies for single-user channels with zero service times studied in \cite{jing-age-online}, after altering them to fit for our two-hop network with non-zero service times. Specifically, let $u_i$ denote the length of the $i$th interval of time that starts with an empty battery at either the source or the relay at time $\ell_{n_i}^-$ for some $n_i\geq1$, i.e., either $\mathcal{E}\left(\ell_{n_i}^-\right)=0$ or $\bar{\mathcal{E}}\left(\ell_{n_i}^-\right)=0$, and ends with both nodes having sufficient energy units to transmit an update at time $\ell_{n_i+u_i}^-$, i.e., with $\mathcal{E}\left(\ell_{n_i+u_i}^-\right)\geq E$ and $\bar{\mathcal{E}}\left(\ell_{n_i+u_i}^-\right)\geq\bar{E}$. Similarly, let $v_j$ denote the length of the $j$th interval of time that starts with both the source and the relay having sufficient energy to transmit an update at time $\ell_{n_j}^-$ for some $n_j\geq1$, and ends when at least one of the two nodes have an empty battery at time $\ell_{n_j+v_j}^-$. With the initial condition $\mathcal{E}=E$ and $\bar{\mathcal{E}}=\bar{E}$, the source and the relay start with sending an update at time $0$. One can then group the scheduled update time slots into $v_1,u_1,v_2,u_2,\dots$, see Fig.~\ref{fig_on_ex_u_v} for a visual example of the above notation. Observe that the shaded area in the figure, representing the area under the age curve between times $0$ and $\ell_1$ does not affect the long term average AoI as it is merely a finite area.

We are interested in the distribution of the $u_i$'s, as it will prove beneficial later in the proof. Let $A$ and $\bar{A}$ be two random variables denoting the number of energy arrivals harvested in between two scheduled update times at the source and the relay, respectively. Since the energy arrival processes at the source and the relay are i.i.d. Poisson with unit rate, it then follows that both $A$ and $\bar{A}$ are i.i.d. Poisson random variables with parameter $\max\left\{1,d+\bar{d}\right\}$. According to the policy, both nodes start with empty batteries at $\ell_{n_i}$ (whichever node has a non-zero amount of energy at $\ell_{n_i}^-$ dumps that amount). Therefore, it holds that\footnote{Without that {\it dumping} condition, analyzing the distribution of the $u_i$'s becomes more complicated, as opposed to the quite simplistic current form.}
\begin{align}
\mathbb{P}\left(u_i=k\right)&=\left(1-\mathbb{P}\left(A\geq1,\bar{A}\geq1\right)\right)^{k-1}\mathbb{P}\left(A\geq1,\bar{A}\geq1\right) \\
&=\left(1-p^2\right)^{k-1}p^2,
\end{align}
where $p\triangleq1-1/e^{\max\left\{1,d+\bar{d}\right\}}$, and the first equality follows from the fact that according to the definition of the update policy, whenever a scheduled update time fails, both nodes reset their batteries. Thus, the only way to have a successful scheduled update after a failed one is that at least one energy arrival occurs at {\it both} nodes. The second equality follows by independence of $A$ and $\bar{A}$. Therefore, $u_i$'s are i.i.d. geometrically distributed random variables with parameter $p^2$, with finite first and second moments. This latter fact will come helpful later in the proof.

Let $K(T)$ denote the number of failed scheduled updates within time $T$. Now let us pick $T$ to be an integer multiple of $\max\left\{1,d+\bar{d}\right\}$, and pick it larger than $\sum_{i=1}^{K(T)}u_i$. Therefore, it holds that
\begin{align}
T=\sum_{i=1}^{K(T)}u_i+N(T)\cdot\max\left\{1,d+\bar{d}\right\}.
\end{align}
We now show that the term $\frac{1}{T}\sum_{i=1}^{K(T)}u_i$, representing the ratio of time in which no updates are transmitted, goes to $0$ a.s. as $T$ grows infinitely large. Toward that end, we first argue that $K(T)\rightarrow\infty$ a.s., which implies by the strong law of large numbers (SLLN), it holds that $\frac{1}{K(T)}\sum_{i=1}^{K(T)}u_i\rightarrow\mathbb{E}\left[u_i\right]$ a.s., the first moment of a geometric random variable with parameter $p^2$, which is finite. We then argue that $K(T)/T\rightarrow0$ a.s., rendering $\frac{K(T)}{T}\frac{1}{K(T)}\sum_{i=1}^{K(T)}u_i\rightarrow0$ a.s. as required. Note that such result implies that the update rate $N(T)/T$ under the best effort uniform update policy with service constraints converges to the upper bound in (\ref{eq_N_T_ub}), which is a necessary step toward showing the achievability of the lower bound on the AoI in Lemma~\ref{thm_on_lb}. The fact that $K(T)\rightarrow\infty$ and $K(T)/T\rightarrow0$ can be shown along the same lines of the proof of \cite[Theorem~1]{jing-online-sensing} (see also the proof of \cite[Theorem~1]{jing-age-online}). We highlight the main difference in what follows. In \cite[Appendix A]{jing-online-sensing}, the amount of energy in the sensor's battery starting at the beginning of an interval $v_j$ until its end is represented by a random walk. It is then shown that such random walk reaches $0$ within a finite time a.s. through a series of indirect steps involving properties of martingales and Poisson processes. Once this is shown, the result follows. In our setting, one can represent the amounts of energy in the source's and the relay's batteries as two i.i.d. random walks, each of which having a finite $0$-hitting time a.s. Since our proposed policy stops if either of the two nodes have an empty battery, i.e., at the minimum of both $0$-hitting times, then this will occur in a finite time a.s. as well. Since, according to our proposed policy, both nodes will have $0$ energy at such hitting time (recall that whichever node has an extra amount of energy it gets rid of it), the rest of the proof in \cite[Appendix A]{jing-online-sensing} directly follows.

Next, let us define the following function denoting the area of the trapezoids that constitute $A_T$:
\begin{align}
h(x)\triangleq\frac{1}{2}\left(x+d+\bar{d}\right)^2-\frac{1}{2}\left(d+\bar{d}\right)^2,
\end{align}
where $x$ denotes the side length of the trapezoid that is on the horizontal axis. Therefore, one can write
\begin{align}
\frac{A_T}{T}&=\frac{\sum_{i=1}^{K(T)}h\left(u_i+\max\left\{1,d+\bar{d}\right\}\right)}{T}+\frac{T-\sum_{i=1}^{K(T)}u_i+1}{T\max\left\{1,d+\bar{d}\right\}}h\left(\max\left\{1,d+\bar{d}\right\}\right) \\
&=\frac{K(T)}{T}\left(\frac{1}{K(T)}\sum_{i=1}^{K(T)}\frac{1}{2}\left(u_i+\max\left\{1,d+\bar{d}\right\}+d+\bar{d}\right)^2-\frac{1}{2}\left(d+\bar{d}\right)^2\right) \nonumber \\
&\quad+\frac{h\left(\max\left\{1,d+\bar{d}\right\}\right)}{\max\left\{1,d+\bar{d}\right\}}\left(1-\frac{K(T)}{T}\frac{1}{K(T)}\sum_{i=1}^{K(T)}u_i+1\right) \\
&=\frac{K(T)}{T}\left(\frac{1}{2K(T)}\sum_{i=1}^{K(T)}u_i^2+\!\!\left(\!\max\left\{1,d+\bar{d}\right\}+d+\bar{d}-\frac{h\left(\max\left\{1,d+\bar{d}\right\}\right)}{\max\left\{1,d+\bar{d}\right\}}\!\right)\!\!\frac{1}{K(T)}\sum_{i=1}^{K(T)}u_i\right. \nonumber \\
&\hspace{1.5in}\frac{h\left(\max\left\{1,d+\bar{d}\right\}\right)}{\max\left\{1,d+\bar{d}\right\}}+\frac{1}{2}\left(\max\left\{1,d+\bar{d}\right\}+d+\bar{d}\right)^2-\frac{1}{2}\left(d+\bar{d}\right)^2\Bigg) \nonumber \\
&\quad+\frac{h\left(\max\left\{1,d+\bar{d}\right\}\right)}{\max\left\{1,d+\bar{d}\right\}}.
\end{align}
Now observe that by SLLN the quantities $\frac{1}{K(T)}\sum_{i=1}^\infty u_i$ and $\frac{1}{K(T)}\sum_{i=1}^\infty u_i^2$ converge, a.s., to the first and second moments of a geometric random variable with parameter $p^2$, respectively, which are both finite quantities. Since $K(T)/T\rightarrow0$ as $T\rightarrow\infty$, it therefore follows that under the best effort uniform update policy with service constraints, the following long term average AoI is achievable:
\begin{align}
\limsup_{T\rightarrow\infty}\frac{1}{T}\mathbb{E}\left[A_T\right]&=\frac{h\left(\max\left\{1,d+\bar{d}\right\}\right)}{\max\left\{1,d+\bar{d}\right\}} \\
&=\frac{\frac{1}{2}\left(\max\left\{1,d+\bar{d}\right\}+d+\bar{d}\right)^2-\frac{1}{2}\left(d+\bar{d}\right)^2}{\max\left\{1,d+\bar{d}\right\}} \\
&=\max\left\{\frac{1}{2}+d+\bar{d}~,~\frac{3}{2}\left(d+\bar{d}\right)\right\}.
\end{align}
This completes the proof.


\end{document}